# Fluid flow vorticity measurement using laser beams with orbital angular momentum

A. Ryabtsev,[1] S. Pouya,[2] A. Safaripour,[2] M. Koochesfahani,[2] and M. Dantus[1]

[1]Department of Chemistry, Michigan State University, East Lansing, Michigan 48824, USA

[2]Department of Mechanical Engineering, Michigan State University, East Lansing, Michigan 48824, USA

We report on the direct measurements of fluid flow vorticity using a spatially shaped beam with a superposition of Laguerre-Gaussian modes that reports on the rotational Doppler shift from microparticles intersecting the beam focus. Experiments are carried out on fluid flows with well-characterized vorticity and the experimental results are found to be in excellent agreement with the expected values even for 200 ms integration times. This method allows for localized real-time determination of vorticity in a fluid flow.

Vorticity is mathematically defined as the curl of the velocity vector, $\Omega = \nabla \times U$, and is physically interpreted as twice the local rotation rate (angular velocity) $\omega$ of a fluid particle, i.e. $\Omega = 2\omega$. It is one of the most dynamically important flow variables and is fundamental to the basic flow physics of many areas of fluid dynamics, including aerodynamics, turbulent flows and chaotic motion. In a turbulent flow, unsteady vortices of various scales and strengths contribute to the chaotic nature of turbulence. The complexity of the problem led Richard Feynman to state that "Turbulence is the most important unsolved problem of classical physics".[1] Even though spatially- and temporally-resolved direct measurement of instantaneous vorticity has been a long-held goal, it has proven elusive to date. Currently in all non-intrusive methods, whether particle-based such as Laser Doppler Velocimetry (LDV) and Particle Image Velocimetry (PIV) [2] or molecular-based as in Molecular Tagging Velocimetry (MTV) [3], vorticity is estimated from a number of velocity field measurements at several points near the point of interest, which then allow computation of the velocity derivatives over space. These methods provide a measurement of vorticity that is spatially averaged over the (small) spatial resolution area of each method. The first direct measurement of vorticity was attempted more than three decades ago by measuring the rotation rate of planar mirrors embedded in 25μm transparent spherical beads that were suspended in a refractive-index-matched liquid.[4] The implementation of this method is very complex and requirement of index matching significantly limits its use and prohibits its application in gas (e.g. air) flows. A capability to directly measure vorticity in fluid flows in a non-intrusive time-resolved manner would greatly impact the field of fluid mechanics.

Direct non-intrusive measurement of vorticity requires a laser-based method that is sensitive to rotational motion. Translational velocities can be measured with laser Doppler velocimetry (LDV) by taking advantage of the (linear) Doppler Effect, which causes a frequency shift when objects move towards or away from a source of light. Analogously, but much less utilized, the Rotational Doppler Effect (RDE) can be used to measure the angular velocity of a rotating object.[5-8] Measuring with RDE requires the use of Laguerre-Gaussian (LG) light beams that possess orbital angular momentum (OAM), a spatial (azimuthal) modulation of the beam phase front. The creation of beams with arbitrary orbital angular momentum $l$, or beams having a superposition of counter-rotating OAM ($\pm l$), requires computer controlled 2D spatial light modulators (SLM) capable of introducing complex phase designs.

The use of LG laser beams with counter-rotating OAM ($\pm l$) to determine the angular speed of rotating objects based on RDE was recently reported by Lavery et al.[6] When the illumination comprises two helically phased beams of opposite values of $l$, their scattering into a common detection mode gives opposite frequency shifts resulting in an intensity modulation of frequency $f_{mod} = 2|l|\omega/2\pi$, where $\omega$ is the angular velocity of the rotating object. Lavery et al.[6] tested this type of setup and were able to measure the angular velocity of a spinning disk. Similar same concepts have been employed to spin and to measure the angular velocity of a microparticle trapped and spinning in an optical trap. [9, 10]

We present here what we believe is the first direct vorticity measurement in a fluid flow based on angular velocity measurement of micron-sized particles free flowing in the fluid using RDE and LG laser beams with OAM. Very small particles faithfully track the fluid flow and, at steady state, they move with the local flow speed and rotate with the local angular velocity of the fluid (or half the local flow vorticity at the particle center) [11]. We demonstrate the technique in a flow field known as solid body rotation or rigid body flow field—for which the angular rotational velocity is uniform and particles carried by the flow also rotate about their center as if they were part of the rigid body. In this type of flow the vorticity is the same everywhere. We present two sets of experiments. In the first, the signal from a group of 6 μm microparticles is integrated to obtain the average fluid rotation rate about the beam optical axis within a 100 μm illumination region, thus obtaining the spatially-averaged vorticity within that region. In the second experiment, the same information is obtained by measuring the angular velocity of a single 100 μm particle in the flow. The latter is the type of transient measurement required to determine vorticity in more complex flows fields.

The experimental setup for measurements of the local flow angular velocity and vorticity is shown in Fig. 1(a). The 488 nm continuous wave beam from an optically pumped semiconductor laser (Genesis MX, Coherent, USA) with initially Gaussian beam profile is expanded by a telescope (L1, L2) and shaped by a two-dimensional liquid crystal on silicon spatial light modulator (LCOS-SLM, Hamamatsu, Japan). The SLM is programmed with a diffraction pattern that introduces the LG spatial modulation and diffracts the spatially shaped beam as shown in Fig. 1(b). The shaped beam possesses the orbital angular momentum corresponding to a superposition of LG ±18 modes, and its far-field intensity profile corresponds to a circular periodic structure with 36 petals (Fig 1(c)). The use of optical angular momentum in the context of the experiments presented here has been reviewed recently.[12] The beam is then focused with long focal length lens L3 and first diffraction order is selected with an aperture. Lens L4 collimates the beam, which after reflection from dichroic mirror (DM) is focused by lens L5 (60mm focal length) into the center of a rotating cylindrical container with the beam optical axis aligned along the rotation axis. The beam diameter at the focus is measured to be about 120 μm and the average power is 12 mW, an intensity that is at least one order of magnitude too weak for causing laser trapping. The container is filled with fluorescent microparticles suspended in a density matched solution of water and glycerin (density about 1.05). Two sets of red fluorescent polymer microspheres (Thermo Fisher Scientific Inc.) are used in these measurements, one with 6 μm diameter (15% variance) and the other 100 μm diameter (7% variance). The container cap is fitted with a thin quartz window that touches the liquid surface at all times to eliminate free surface effects. The angular velocity $\omega$ of the container is controlled by an optically encoded motor (3501 Optical Chopper, New Focus, USA) rotating at frequency $f$ and angular velocity $\omega = 2\pi f$. Measurements were done after the container was spun for a few minutes to ensure a steady state rotation flow field had been established. The resulting flow field is devoid of any secondary flow and is precisely characterized by the solid-body rotation velocity field $U = r \times \omega$ and its spatially uniform vorticity field $\Omega = 2\omega$.

Epi-directional fluorescent light from the irradiated particles is collected with lens L6 and is focused onto a photodiode. A small diameter pinhole is set before photodiode in order to spatially filter out signal from outside of focal volume in fluid. The intensity modulated signal from the fluorescent particles is recorded at 10 kHz sampling rate and spectrally analyzed. Earlier efforts from our group to detect vorticity based on back scatter resulted in very poor signal to noise ratios due to multiple sources of scatter.[13] The use of epi directional detection and the use of fluorescent particles, in combination with clean OAM shaped laser excitation, allows us to reject scattered light from the rotating surfaces of the container and guarantees the measured signal originates from within the rotating body of fluid.

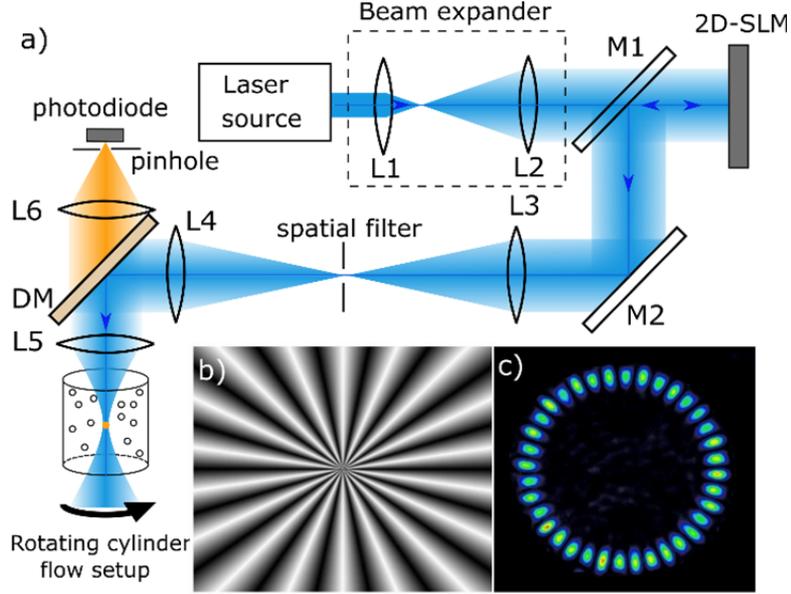

FIG. 1. (a)Experimental setup. L1-L6, lenses; M1-M2, mirrors; DM, dichroic mirror. (b) Diffraction pattern displayed on 2D White color corresponds to 0 phase shift while black corresponds to 2π phase shift with 256 steps in between. (c) Resulting beam structure used to illuminate particles in fluid flow.

For the measurements presented, the LG laser beam has an OAM with $l = \pm 18$, resulting in 36 bright features (petals). Scattering from objects rotating at angular velocity $\omega$ (or rotation frequency $f$) leads to intensity modulation at frequency $f_{mod} = 36 \omega/2\pi = 36 f$. The first set of data in Figure 2 shows the measurement with 6 μm fluorescent particles. In this case, by measuring the rotation rate of an ensemble of particles within the ≈ 100 μm beam diameter we obtain the averaged fluid rotation rate within that region. Figure 2 (a) shows examples of intensity modulation of collected (AC-coupled) signal for four different prescribed rotation frequencies of the cylindrical container. Fourier transforms of each signal provides the spectral information in Fig. 2 (b) using a short data record of about 200 ms in length.

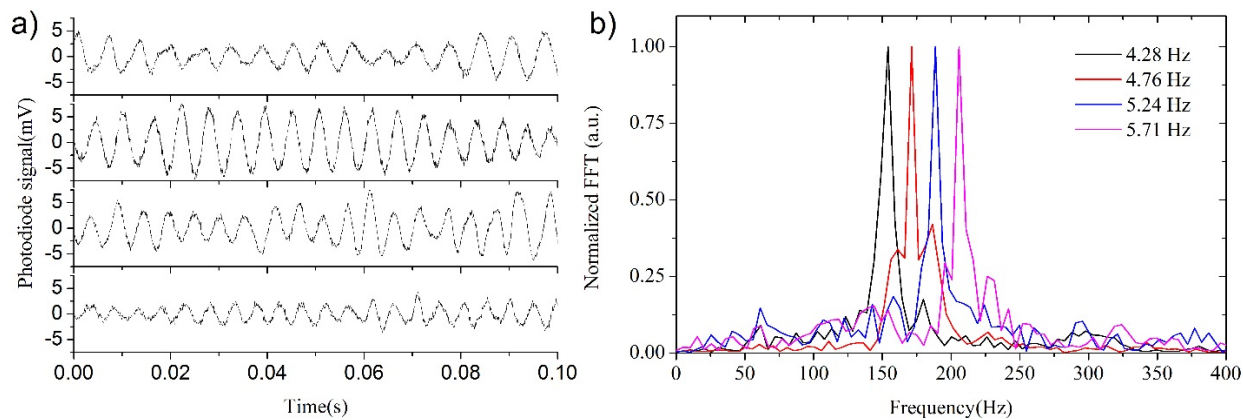

FIG. 2. (a) 100 ms long time series of collected signal for four different rotation frequencies of the container: $f$ = 4.28 Hz, 4.76 Hz, 5.24 Hz and 5.71 Hz. (b) Power spectrum of the signals in (a) (200 ms data record).

From the spectral peaks in Figure 2 (b) we measure the modulation frequencies for the four cases to be 154.16±5 Hz, 171.37±5 Hz, 188.58±5 Hz, and 205.76±5 Hz. These values correspond to the measured fluid rotation rates of 4.28±0.14 Hz, 4.76±0.14 Hz, 5.24±0.14 Hz, and 5.72±0.14 Hz, respectively. These values are in excellent agreement with the prescribed rotation frequencies of the rotating fluid container. Accuracy of these measurements is limited by FFT resolution divided by two-times the OAM. For a 200 ms data record and OAM with $l = \pm 18$ we obtain 5/36 = 0.14 Hz. Given the steady flow field in this experiment, one can improve the measurement accuracy, if desired, by increasing the length of the data record for FFT analysis or increasing the beams OAM. For arbitrary flows, one could speed up data acquisition to 20 ms, reducing accuracy to 1.4 Hz.

The second set of experiments was carried on with larger 100 μm particles with low particle density in solution to ensure single particle measurement within the ≈ 100 μm beam diameter. This was confirmed by visually observing the single particle presence in the focal volume of structured laser beam based on its intensity time series during data collection. FFT analysis for two different prescribed rotation frequencies of the cylindrical container, $f$ = 4.28 Hz and 4.76 Hz, is presented in Fig. 3. The peaks indicate modulation frequencies of 154.08±5 Hz, and 170.10±5 Hz for these two cases. The corresponding values of the measured fluid rotation rates are 4.28±0.14 Hz, 4.73±0.14 Hz, which are again in excellent agreement with the imposed rotation frequencies of the rotating fluid container.

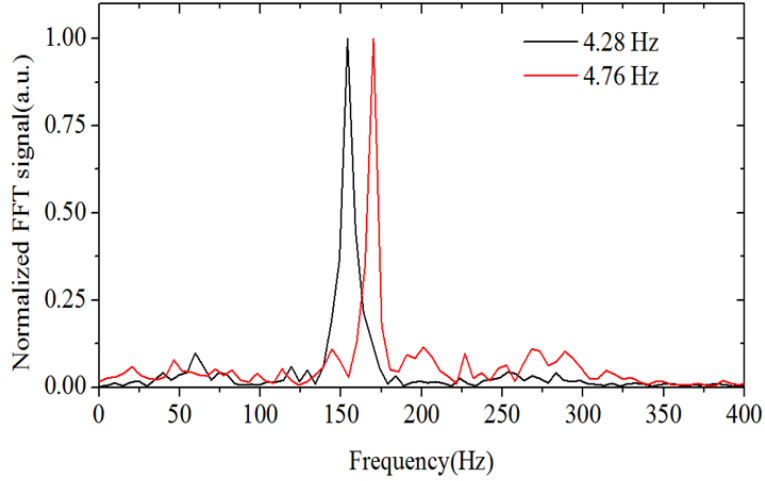

FIG. 3. Power spectrum of signal for a single 100μm particle in solution for two different rotation frequencies of the cylindrical container: $f$ = 4.28 Hz, and 4.76 Hz (200 ms data record).

The solid body rotation flow field was selected for these proof-of-concept experiments because it is relatively simple to create and has well-characterized velocity and vorticity fields. When the liquid-filled cylindrical container initially at rest starts to spin, the fluid layer near the moving wall starts to move with the cylinder due to the no-slip viscous boundary condition at the wall. The motion is then propagated throughout the container by viscous shear until the entire body of liquid rotates at the same speed of the container. The final steady state velocity field is that of solid body rotation with vorticity that is constant in time and uniform in space, with axis parallel to the axis of rotation of cylinder and magnitude equal to twice the cylinder angular velocity. While we have demonstrated here the idea of vorticity measurement using laser beams with OAM in a steady flow environment, clearly most exciting applications would be in unsteady flows. For micro particles in Stokes flow regime, particle rotation time can be estimated from $\tau = \rho_p d^2 / 60\mu$, where $\rho_p$ and $d$ are the particle density and diameter, and $\mu$ is the fluid viscosity.[14] For 100 μm particles like those used in our experiment the rotation time is about 100 μsec. Therefore, unsteady vorticity measurements are feasible and could be obtained by acquiring shorter record lengths of data. Because of the quadratic dependence of particle rotation time on diameter, one can select the appropriate particle size to ensure a response time that is faster than the flow-fluctuation time scale. We performed this type of measurement but under steady flow conditions. The time invariant spectral peak (200ms FFT window) in Figure 4 is obtained from a 100 μm particle that lingers within laser illumination at the axis of the rotating container while spinning with the fluid rotation rate. Measuring the spectral peak based on the 40 s data record yields a modulation frequency of 171.5±0.025 Hz, or particle/fluid rotation rate of 4.76 Hz, in perfect agreement with the imposed rotation frequency of the fluid container.

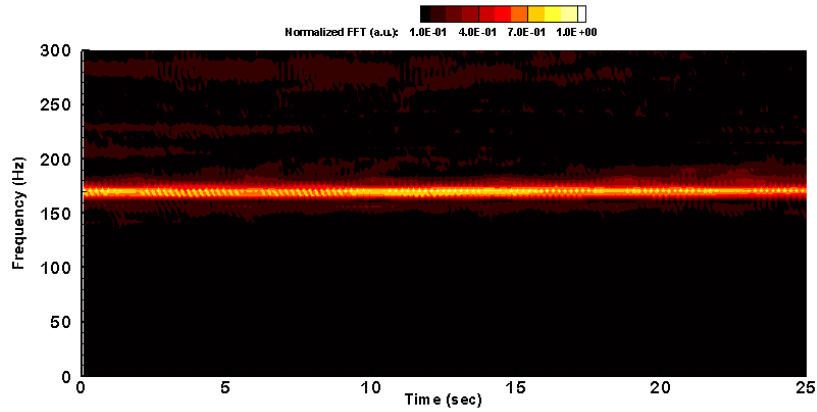

FIG. 4. FFT map of signal for a single 100 μm particle in solution over time ($f$ = 4.76 Hz)

While the experiments we have reported here represent the extension of the work of Lavery et al. [6] to the field of fluid dynamics, there are certain differences between the two as well. In the latter, the scattering signal originates from the planar surface of a spinning disk. In ours, measurements are carried out within the body of the fluid and the scattering signal is from a finite volume inside the fluid. For measurements with high spatial resolution, this scattering volume needs to be localized to a small region. In the current experiments, this was achieved by focusing the laser beam to about 100 μm diameter inside the liquid container.

We have presented the first direct and localized non-intrusive measurement of vorticity in a fluid flow using the Rotational Doppler Effect and Laguerre-Gaussian spatially modulated light beams that possess orbital angular momentum. The approach has been demonstrated in the flow field of solid body rotation where the flow vorticity is known precisely. In one experiment, measurements with a group of 6 μm microparticles is used to obtain the average fluid rotation rate about the beam optical axis within the 100 μm illumination region, and therefore, the spatially-averaged vorticity within. In another experiment, the same information is obtained by measuring the angular velocity of a single 100 μm particle in the laser beam. In both experiments, the measured results are in excellent agreement with those expected from the prescribed rotation frequencies of the rotating fluid container.

Although, the technique is demonstrated here in a simple flow where vorticity is uniform and steady, the approach holds great promise for unsteady flows with spatially varying vorticity field. We plan to explore extensions of this measurement technique to more complex flow environments.

**ACKNOWLEDGMENTS**

The work was funded by the Air Force Office of Scientific Research (AFOSR) FA9550-14-1-0312 (program manager Dr. Douglas Smith). The authors gratefully acknowledge seed funding for this work from the Michigan State University Office of the Vice President of Research and Graduate Studies, Dr. Steven Hsu.